\documentclass[twocolumn,showpacs,preprintnumbers,amsmath,amssymb]{revtex4}

\usepackage{graphics}

\usepackage{dcolumn}
\usepackage{bm}

\begin{document}

\draft


\title{Analysis of Low Energy Pion Spectra}
\author{Suk Choi and Kang Seog Lee\footnote{kslee@chonnam.ac.kr} }
\address{Department of Physics, Chonnam National University,
Gwangju 500-757, Korea}

\date{Apr. 11, 2007}

\begin{abstract}
The transverse mass spectra and the rapidity distributions of
$\pi^+$ and $\pi^-$ in Au-Au collisions at 2, 4, 6, and 8
GeV$\cdot$A by E895 collaboration are fitted using an elliptically
expanding fireball model with the contribution from the resonance
decays and the final state Coulomb interaction. The ratio of the
total number of produced $\pi^-$ and $\pi^+$ is used to fit the
data. The resulting freeze-out temperature is rather low($T_f <
60$ MeV) with large transverse flow and thus resonance
contribution is very small. The difference in the shape of $m_t$
spectra of the oppositely charged pions are found to be due to the
Coulomb interaction of the pions.

\end{abstract}

\pacs{24.10.Pa,25.75.-q }

\maketitle

%


Pion production just above the threshold energy is quite different
from that at very high energies such as RHIC energy since the
ratio of $\pi^-$ to $\pi^+$ at very high energies is one which is
not the case at low energies. At just above the threshold energy,
pions are produced through the production of $\Delta$ resonances
and counting all the possible channels of $\Delta$ decay the
difference in the composition of isospins in the colliding nuclei
appears as the difference in the numbers of the two oppositely
charged pions\cite{verWest,wagner,stock}, whereas at high energies
many channels producing pions are open and small asymmetry in the
initial isospin does not matter.

Other features of the pion spectra at low energies
are\cite{stock,verWest,klay,hong2,hong,rafelski,wagner}: (1)The
transverse momentum spectra both of the $\pi^-$ and $\pi^+$ seem
to have two temperatures. Usually the low temperature component in
the low momentum region is attributed to the pions decayed from
resonances, especially the delta resonance, while the higher
temperature component in the mid-momentum region is the thermal
ones. (2) Transverse momentum spectra of $\pi^-$ and $\pi^+$ at
very small momentum are different in the sense that the $\pi^+$
spectra is convex in its shape while the $\pi^-$ spectra does not
show this behavoir. This difference in low momentum region is due
to the Coulomb effect. The hadronic matter formed during the
collision has charge which comes from the initially colliding two
nuclei and thus the thermal pions escaping from the system
experience the Coulomb interaction. The Coulomb interaction of
$\pi^-$ may bend the spectrum in the low momentum region upward
and thus it is hard to disentangle the contribution from the delta
resonance and the Coulomb interaction in the low momentum region.
(3) Width of the rapidity spectra of $\pi^-$ and $\pi^+$ are much
wider than those from the isotropically expanding thermal model.
The wide width may either come from partly transparent nature of
the collision dynamics or the ellipsoidal expansion geometry. In
order to fit large rapidity width using expanding fireball model
one usually needs large longitudinal expansion velocity.

Even though all those features mentioned above are not new,
calculations with all those features put in together comparing
each contributions in detail is hard to find. There are claims
that the properties of $\Delta$ resonance are modified inside the
hadronic matter formed even at this low
energy\cite{hong,rafelski}. In order to draw any conclusion, one
should have a model which can explain all of the features above
mentioned. Lacking one or two features in the calculations, the
result may not be conclusive.

In this paper, we analyze the pion spectra in Au+Au collisions at
2, 4, 6, and 8 A$\cdot$GeV measured by the E895
collaboration\cite{klay} using the expanding
fireball\cite{lee,sollfrank,dobbler} with the resonance
contribution\cite{sollfrank} and the final state Coulomb
interaction\cite{gyulassy,heiselberg,kapusta,wagner}. The geometry
of the expansion used is ellipsoidal\cite{dobbler} and can be
varied to sphere and cylinder, by taking the transverse size $R$
as a function of the longitudinal coordinate, $z$.

At just above the threshold energy, pions are produced through the
production of $\Delta$ resonances and their subsequent decays. The
ratio of $\pi^-$ and $\pi^+$ is given from the initial isospin
conservation as $\frac{\pi^-}{\pi^+} = \frac{5N^2+NZ}{5Z^2+NZ}\sim
1.94$ for Au+Au collisions\cite{verWest,wagner,stock} . Hence it
is expected that at low beam energies near 2 GeV$\cdot$A in Au+Au
collisions, the ratio is near 1.94 and then as the beam energy is
increased the ratio will decrease eventually to one. In the
present calculation, the ratio of normalization constants for
$\pi^-$ and $\pi^+$, $R$ is taken as a fit parameter in order to
investigate the beam energy dependence of $R$.


We assume that once the pions are produced they rescatter among
themselves and thermalize before they decouple from the system.
Hence we assume thermalization of the pionic matter. However, the
total number of negatively and positively charged pions are not in
chemical equilibrium and the ratio is governed by the isospin
asymmetry of the initially colliding nuclei. We keep the ratio as
a fitting parameter and want to compare with the expected value of
1.94 near threshold of the pion production.

After the formation of a pionic fireball, it expands and cools
down until freeze-out when the particle production is described
from the formalism of Cooper-Frye\cite{cooper}. For the
equilibrium distribution function we use the Lorentz-boosted
Boltzmann distribution function, where $u_\mu$ is the expansion
velocity the space-time of the system.

\begin{equation}
E\frac{d^3N}{d^3p} = \frac{g}{(2\pi)^3} \int_{\Sigma_f} p^\mu
d\sigma_\mu (x) f(x,p)
\end{equation}
where
\begin{equation}
 f(x,p) = \exp(-\frac{p_\nu u_\nu (x) - \mu }{T})
\end{equation}

\begin{table}
\caption{Fitted values for each parameters.}
\begin{tabular} {|c|c|c|c|c|c|c|c|}\hline
   $E_{beam}$ & V& $\eta_m$ & $\rho_0$ & T & P$_c$ & $\pi^-/\pi^+$
   & $\chi^2$ /n \\
   (GeV)& ($\times 10^5$) & & & GeV  & GeV/c & &  \\   \hline

2&1.41& 1.12 & 0.88 & 46 & 25 &  1.96& 1.3 \\ \hline

4& 0.93& 1.32 & 0.92 & 57 & 24 &  1.95& 2.9 \\ \hline

6& 1.44& 1.50 & 1.11 & 54 & 18 &  1.40 &2.4 \\ \hline

8& 1.62& 1.58 & 1.12 & 55 & 15 &  1.38 &1.8 \\ \hline

\end{tabular}
\end{table}

For the geometry of the expanding fireball\cite{dobbler}, we
assume azimuthal symmetry and further assume boost-invariance
collective dynamics along the longitudinal
direction\cite{bjorken}. In this case it is convenient to use as
the coordinates the longitudinal proper time $\tau = \sqrt{t^2
-z^2}$, space-time rapidity $\eta = \tanh^{-1}(z/t)$ and the
transverse coordinate $r_\perp$. Then the 4-velocity of expansion
can be expressed as $u^\mu (x)=\gamma (1, v_\perp (x) \bold{e}_r ,
v_z (x))$ where $v_z (\tau, r_\perp , \eta ) = \tanh \eta$, which
is the result of the longitudinal boost-invariance. In the
transverse direction we take a linear flow rapidity profile,
$\tanh^{-1} v_\perp = \rho(\eta) (r_\perp /R_0 )$ where $R_0$ is
the transverse radius at midrapidity. Here one takes $\rho(\eta) =
\rho_0 \sqrt{1-(\eta^2 /\eta_{max} ^2 )}$ for the elliptic
geometry and a constant value of $\rho$, {\it i.e.} $\rho (\eta) =
\rho_0$ for the cylindrical geometry. As is the same for the SPS
energy by Dobbler {\it et. al.}, the elliptic case fits the pion
spectra a little better.

As pions escape from the system at freeze-out, they experience the
Coulomb interaction with the charge of the system which are mainly
due to the initial protons in the colliding nuclei. The Coulomb
effect on the particle spectra are studied in detail in
refs.\cite{gyulassy,kapusta,heiselberg} for the static and
dynamical cases. Here due to the low beam energy we restrict
ourselves only to the static case. The system is expanding rapidly
in the longitudinal direction and the change in the longitudinal
momentum is negligible. Only the transverse momentum of the
charged particles will be shifted by an average
amount\cite{heiselberg}
\begin{equation}
p_c = \Delta p_{\perp} \sim 2e^2 \frac{dN^{ch}}{dy} \frac{1}{R_f},
\end{equation}
where $R_f$ is the transverse radius of the system at freeze-out.
Due to the lack of detailed knowledge, $p_c$ is taken as a fit
parameter in the present calculation. And in this way the beam
energy dependence of $p_c$ can be studied.

Since the transverse momentum of the escaping thermal pions are
shifted by the amount $p_c$, {\it i.e.} $p_t = p_{t,0} \pm p_c $.
the invariant cross section can be written in terms of the
unshifted momentum ($p_{t,0}, y_0$).

 \begin{equation}
E \frac{d^3N}{dp^3} = (E \frac{d^3N}{dp^3 })_0 (\frac{dp^3 }{E})_0
(\frac{E}{dp^3 }).
 \end{equation}
 where $(E \frac{d^3N}{dp^3 })_0$ is the unshifted invariant
 cross section.
 Integrating over the rapidity $y$ one gets
 the equation for the transverse momentum spectra,
\begin{equation}
 \frac{dN}{p_t dp_t} = \int dy \frac{d^2N}{p_{t,0} dp_{t,0} dy}
 \frac{p_{t,0}}{p_t}.
 \end{equation}
and the rapidity spectra is obtained by integrating over the
transverse mass.

\begin{figure}
  \begin{center}
       \resizebox{60mm}{!}{\includegraphics{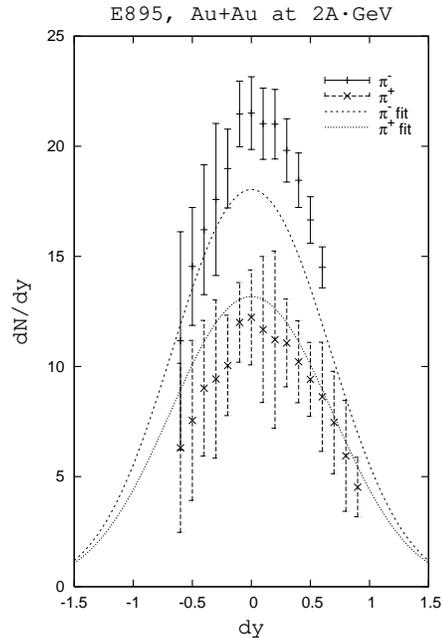}}
  \end{center}
   \caption{ Rapidity distribution of $\pi^-$ and $\pi^+$ in
   Au+Au collision at 2 A$\cdot$GeV by E895 collaboration\cite{klay}.}
  \label{fig:2gevy}
\end{figure}

\begin{equation}
 \frac{dN}{dy} = \int m_{t,0} dm_{t,0} (\frac{d^2N}{p_{t,0} dp_{t,0}})
 \frac{p_{t,0}}{p_t}.
 \end{equation}

Finally one has to add the contribution from resonance
decay\cite{sollfrank} to both the transverse spectrum and rapidity
distribution. Here we assume that the resonances decay far outside
the system and the Coulomb interaction of the pions decayed from
the resonances is neglected.

\begin{figure}
  \begin{center}
     \begin{tabular}{cc}
       \resizebox{40mm}{!}{\includegraphics{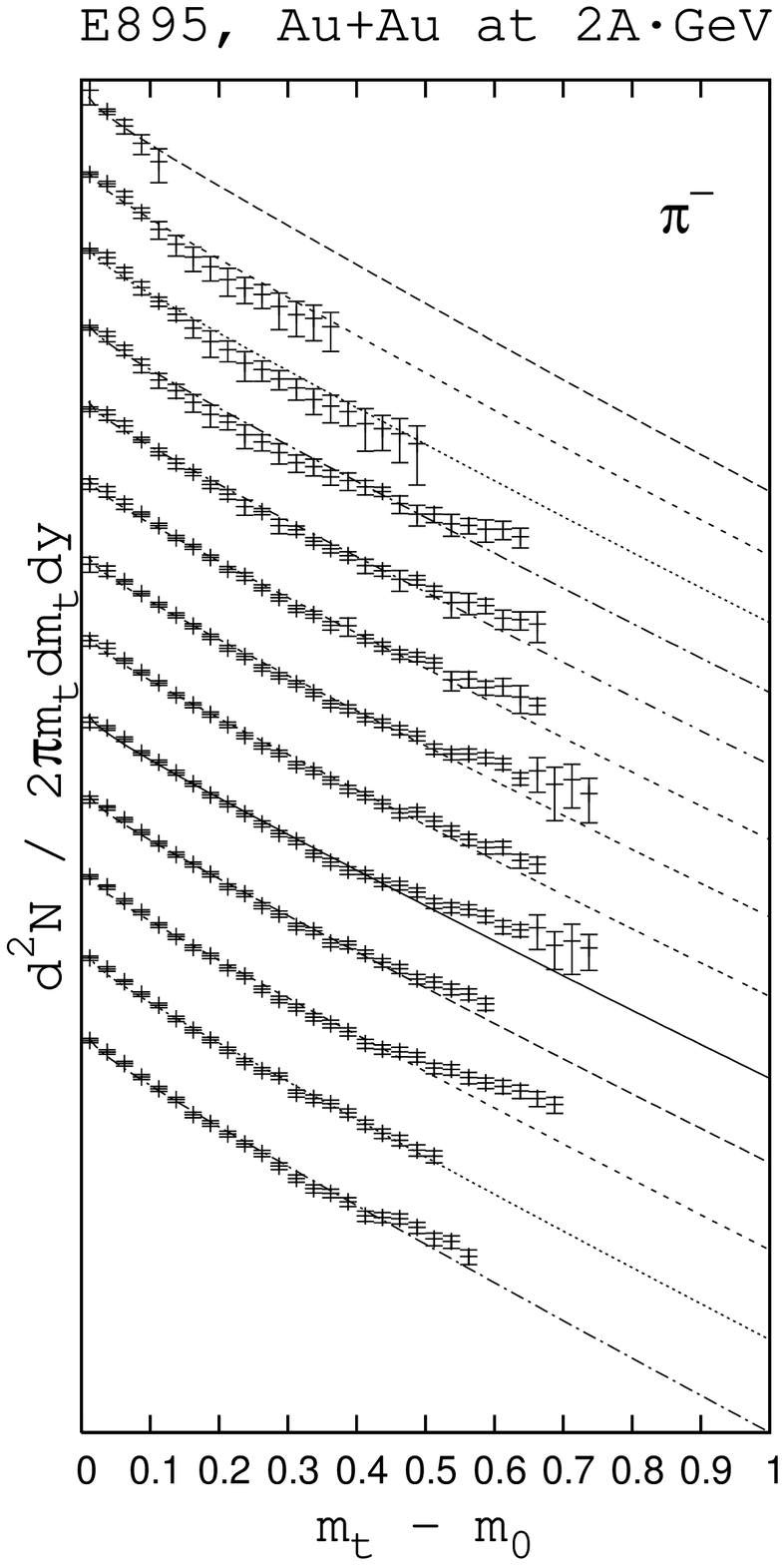}}&
       \resizebox{40mm}{!}{\includegraphics{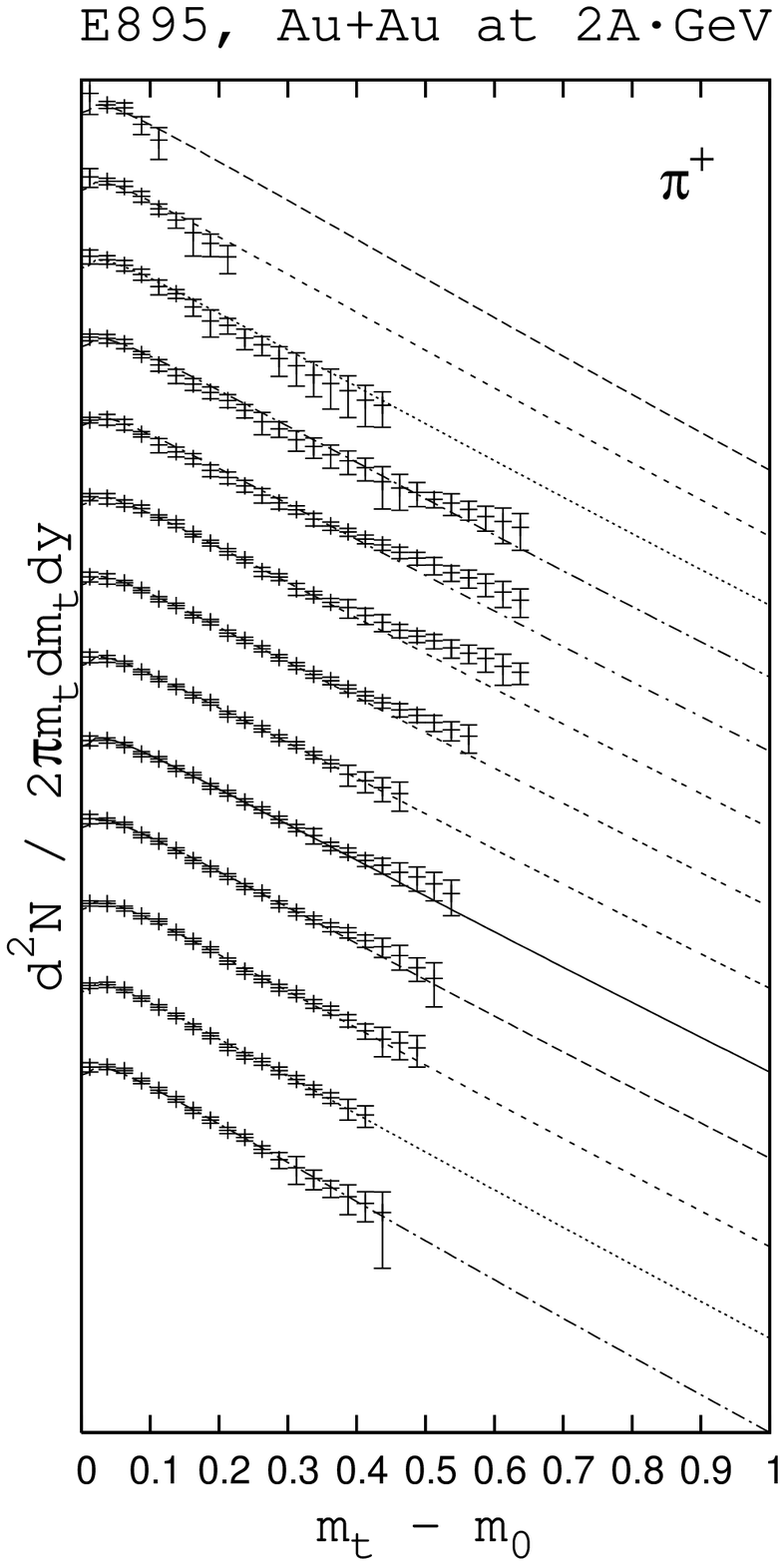}} \\
     \end{tabular}
  \end{center}
   \caption{ $\pi^-$ transverse mass spectrum for each rapidity
  bin($\Delta y = 1.0$) measured by E895 collaboration\cite{klay}in Au+Au
  collisions
  at 2 A$\cdot$GeV. Data on the top line is for the rapidity bin
 $-0.65 <y<-0.55$ and the next one is for $-0.55 <y<-0.45$ scaled by 0.1, etc. }

 \label{fig:2gevpimmt}
\end{figure}

The fitted values for the parameters are tabulated in the Tab.~1
and the results of the fitting are shown in Figs.1-2 for 2
A$\cdot$GeV.
 The fitted value for the ratio $R = \pi^- /\pi^+$
is close to 1.94 at 2 and 4 A$\cdot$GeV as expected and decreases
to 1.38 at 8 A$\cdot$GeV, which eventually becomes 1 at higher
energies such as RHIC energies. In other words, at this very low
energy pion isospin is not in chemical equilibrium.

The freeze-out temperature is rather small, $T_f < 60$ MeV and the
expansion velocities in both the longitudinal and transverse
direction are quite large. The large longitudinal expansion
velocity($> 0.8$c) is needed to fit the large width of the
rapidity distribution. The low freeze-out temperature together
with the large transverse expansion fits the transverse spectra of
pions quite nicely. If not for the large expansion velocity, one
usually gets much larger freeze-out temperature($T_f > 80$ MeV.

Since the freeze-out temperature is small, there are very few
resonances at freeze-out, especially $\Delta$, and thus the
resonance contribution is negligible. This is reasonable since at
this low energy near the pion threshold energy, production of
particles with mass larger than pions is rare and their
contribution to the pion spectrum is negligible.

The pion transverse momentum spectra looks like that there are two
slopes; one for small momentum region and another for the higher
momentum region near the pion mass. The smaller slope at lower
transverse momentum is usually attributed to the pions from the
resonance decay, especially from the $\Delta$
decay\cite{stock,hong}. However, present calculation shows that
this is not the case at low beam energies.

The shape of the transverse mass spectra of $\pi^-$ and $\pi^+$
are different especially in the small mass region. As the
transverse mass $m_t$ decreases, the $m_t $ spectrum of $\pi^-$
increases sharply while that of $\pi^+$ saturates showing the
convex shape. This difference is due to the Coulomb interaction of
pions leaving the system which has the charge from the initially
bombarding nucleons.  The change in the transverse momentum due to
the Coulomb  interaction decreases from 25 GeV/c at 2 A$\cdot$GeV
to 15 GeV/c at 8  A$\cdot$GeV. This behavior can be understood
from the increase of the screening  effect since the number of
charged pions increase at higher energies. At very high energies
such as RHIC energies, the momentum change from the Coulomb
interaction will be small.

 The emerging picture of pion production at low energy is that the pions
 are produced through the intermediate $\Delta$ formation and thus
 they are not in chemical equilibrium in isospin. They make
 collisions and thermalize to form a fireball which expands and cools
 until the freeze-out. Since the fireball has charge which is from
 the initially colliding nucleus, the pions leaving the system
 experiences the Coulomb interaction which makes the difference of
 the $m_t$ spectra of the two oppositely charged pions.

\acknowledgments
This work is financially supported by Chonnam
National University and the post-BK21 program. We wish to
acknowledge U.W. Heinz for providing the program and useful
discussions.


\end{document}